\documentclass[preprint]{aastex63}
\usepackage{amsmath}
\usepackage{amssymb}
\usepackage{graphicx}
\usepackage{color}
\usepackage{xcolor}
\usepackage{xfrac}
\usepackage{animate}
\usepackage[caption=false]{subfig}
\usepackage{rotating}
\usepackage{hyperref}
\usepackage[utf8]{inputenc}

\begin{document}

\title{A Wave-Corrected Assessment of the Local Midplane}

\author{Ziyuan Yin}
\affiliation{Department of Physics, 
Colorado College, Colorado Springs, CO 80903}
\affiliation{Department of Physics, 
Emory University, Atlanta, GA 30322}
\author[0000-0002-9785-914X]{Austin Hinkel}
\affiliation{Department of Physics and Astronomy, Thomas More University, Crestview Hills, KY 41017}
\affiliation{Department of Physics, 
Colorado College, Colorado Springs, CO 80903}

\date{\today}

\begin{abstract}
    As the number of known Galactic structures mounts thanks to the Gaia Space Telescope, it is now pertinent to study methods for disentangling structures occupying the same regions of the Milky Way.  Indeed, understanding the precise form of each individual structure and the interactions between structures may aid in understanding their origins and chronology.  Moreover, accounting for known structures allows one to probe still finer Galactic structure.  In order to demonstrate, we have developed an Odd Low-Pass Filter (OLPF) which removes smaller, odd-parity structures like the vertical waves, and use the filtered data to examine the location of the Galaxy's mid-plane.  We find that the radial wave identified by Xu et al. (2015) continues inward to at least the Sun's location, with an amplitude that decreases towards the inner, denser parts of the disk, consistent with a simple, qualitative simulation.  Additionally, we employ the OLPF results to determine the solar offset, $z_{\odot}$, with smaller structures filtered out.  We find that $z_{\odot} = 34.2 \pm 0.3$ pc.
\end{abstract}

%
%
\section{Introduction} \label{sec:Intro}

With the first light of the Gaia Space Telescope \citep{brown2016gaia, prusti2016gaia, lindegren2018gaia}, superb astrometric data has made it increasingly clear that the Milky Way Galaxy is not in equilibrium \citep[e.g.]{gardner2021milky, hunt2022milky}.  This disequilibrium, apparent in spiral phase-space structures \citep{antoja2018dynamically}, vertical waves \citep{widrow2012galactoseismology, yanny2013stellar, bennett2018vertical}, warping \citep[e.g.]{levine2006vertical, skowron2019three, chen2019intuitive}, and corrugated waves proposed by \citet{bland2021galactic} to name a few, is a direct challenge to the commonly employed assumptions of Galactic dynamics \citep{binney2008GD}.  Indeed, the Galaxy is not in steady state \citep{GHY20}, does not exhibit axial symmetry \citep{HGY20}, and its stars are not entirely uncorrelated \citep{hinkel2023two}.  

Such findings, while challenging the existing theoretical framework, offer unprecedented opportunities to disentangle past events in the Galaxy's history \citep{gardner2021milky}, as different interactions imprint distinct structures within the Milky Way \citep{hunt2022multiple}.  In other words, {\it the distinct pattern of how} particular assumptions of symmetry or equilibrium are broken is important in determining possible causes \cite{hinkel2021axial}.  Such a disentangling effort is critical for understanding the individual effects in detail as well as in the search for additional structure.

The picture near the solar neighborhood, however, may be even murkier.  A number of local structures near the Sun have been found, including potential resonances of the Galactic bar \citep[e.g.]{dehnen2000effect, hunt2018outer, hinkel2020axial, trick2021identifying}, the Radcliffe Wave \citep{alves2020galactic}, as well as a litany of other structures.

The cacophony of these effects add up so as to frustrate any measurement of a single effect.  
To make matters worse, interactions between observational constraints and known structures can tend to bias other measurements.  
That is, cutting on (e.g.) Galactic latitude, $b$, may influence which portion of the vertical wave \citep{widrow2012galactoseismology} one observes at different distances from the Sun.  
This can conceivably skew measurements of the location of the Galaxy's mid-plane due to the odd parity of the vertical waves.

In this paper, we detail a method for disentangling effects by exploiting their distinct symmetries and scales.  
In what follows we briefly outline the underlying theory of our method (Section~\ref{sec:Theory}) and develop this method for precision measurements of Galactic structure
(Section~\ref{sec:Method}).  
Finally, we close with the results of our analysis (Section~\ref{sec:Analysis}) and discuss the possible ramifications and origins of the structures we have found (Section~\ref{sec:Conclusions}).

%
%
\section{Theory} \label{sec:Theory}

Supposing (incorrectly) for the moment that the Milky Way is in equilibrium, the vertical stellar density profile of the Milky Way would be symmetric about the mid-plane, $z = 0$ \citep{an2017reflection}. Indeed, the Milky Way has been described by various, simplistic models, which include contributions from both the disk and other populations like the stellar halo which all exhibit reflection symmetry about the mid-plane \citep[e.g.]{spitzer1942dynamics, miyamoto1975disk, navarro1997universal}.  In other words, the vertical structure described by an equilibrium model would take the form:
\begin{equation}
    n_{\rm eq}(z) = \sum_{j=0}^{\inf} a_{2j} z^{2j} = a_0 + a_2 z^2 + a_4 z^4 + ...,
    \label{Eq:Equilibrium}
\end{equation}
where $z$ is the direction perpendicular to the Galactic plane, and $a_{2j}$ are coefficients in the expansion of the equilibrium model.  
In this case, the mid-plane of the Galaxy is located at the position of maximal density.

However, the Milky Way is clearly in disequilibrium \citep[e.g.]{antoja2018dynamically} and may exhibit over- and under- densities, symmetry breaking, and other effects which can skew an assessment of the local mid-plane due to odd-parity terms not considered in Eq.~\ref{Eq:Equilibrium}.  To insulate a determination of the vertical location of the mid-plane against these effects, we start by considering the average value of $z$ for all stars in a given sector of the Milky Way:
\begin{equation}
    \bar{z} = \frac{\int_{-z_{\rm \ell}}^{z_{\rm \ell}} z n(z) dz}{\int_{-z_{\rm \ell}}^{z_{\rm \ell}} n(z) dz},
    \label{Eq:meanZ}
\end{equation}
where $z_{\rm \ell}$ represents some bounding value of $z$ measured from the nominal, zero-latitude ($b = 0^{\circ}$) plane.  Notice an even-parity distribution will yield $\bar{z} = 0$.  (As $b = 0^{\circ}$ is measured from the Sun and neglecting other effects for the moment, the true mid-plane will be offset from $z$ by an amount $z_\odot$, the Sun's height above the mid-plane.)

There are many effects in the Galaxy which appear to factor into the value of $\bar{z}$, and thus an assessment of $z_\odot$, as motivated in Section~\ref{sec:Intro}.  To this end, it is useful to understand the parity of the various other effects, and so we introduce a general set of basis functions, $\{e_n(z)\}$ and $\{g_n(z)\}$, such that all $e_{n}(z)$ have even parity, while all $g_{n}(z)$ are odd.  In this manner, we can describe the various effects as weighted sums of some subset of the $\{e_n\}$ and $\{g_n\}$, as generalized for some arbitrary function, $h(z)$, in Eq.~\ref{Eq:basisExample}:
\begin{equation}
    h(z) = \sum_{n=0}^{\infty} \left(a_n  e_n(z) + b_n  g_n(z)\right),
    \label{Eq:basisExample}
\end{equation}
where the $a_n$ and $b_n$ represent the coefficients in the expansion of $h$ in the $\{e_n, g_n\}$ basis set.  (In our particular case, we have chosen the Fourier basis.)

Regardless of the choice of basis functions, the explicit even and odd parity terms allow for various simplifications in the computation of $\bar{z}$.  To see why, suppose we have a vertical distribution function, $n_{\rm eq}(z)$, describing a Milky Way in equilibrium as in Eq.~\ref{Eq:Equilibrium}.  As Eq.~\ref{Eq:Equilibrium} is expressly even, we may write it as:
\begin{equation}
    n_{\rm eq}(z) = \sum_{n=0}^{\infty} a_{{\rm eq},{n}}  e_n(z).
    \label{Eq:n_eq}
\end{equation}

Moreover, suppose that the various structural effects discussed above factor into the vertical distribution of stars as small corrections, so that we may add terms to the equilibrium condition as a small perturbation.  While we do not {\it a priori} know the form of the perturbations, $n_{\rm pert}(z)$, we do know that it can contain non-zero, odd parity terms, which can act to shift the vertical location of the mid-plane.  Further, there are various known effects with known parities and scales.  For example, symmetric breathing modes \cite[e.g.]{ghosh2022age} would possess even parity, while the vertical waves \citep{widrow2012galactoseismology} would possess odd parity.  In the case of the latter, the waves have scales on the order of half a kiloparsec \citep{bennett2018vertical}.

Symmetric perturbations like the breathing modes mentioned above will not shift the location of the Milky Way mid-plane.  Effects with odd-parity terms will shift the mid-plane\footnote{\citet{bennett2018vertical} calculate a wave-corrected assessment of $z_\odot$ based solely on the vertical star count asymmetry in a cylinder around the sun of radius 250 pc, whereas this work allows for the morphology of the vertical wave to vary between regions, as motivated by \citet{hinkel2023two}.}, and so we define an Odd Low Pass Filter (OLPF) such that we filter out all odd (imaginary) terms in the Fast Fourier Transform of the vertical number counts, $\psi(\Tilde{\nu})$, exceeding some wavenumber (i.e., with correspondingly small scales).  More formally, the OLPF, $F(\Tilde{\nu})$, as a function of the linear wavenumber $\Tilde{\nu}$ transforms $\psi(\Tilde{\nu})$ in the following way:
\begin{equation}
        F(\Tilde{\nu}) \psi(\Tilde{\nu})=
        \left\{ \begin{array}{ll}
            \psi(\Tilde{\nu}) & \Tilde{\nu} \leq \Tilde{\nu}_{\rm lim} \\
            \Re\left(\psi(\Tilde{\nu})\right) & \Tilde{\nu} > \Tilde{\nu}_{\rm lim}
        \end{array} \right. ,
        \label{Eq:OLPF}
\end{equation}
where $\Re(x)$ returns the real part of $x$.
Upon filtering the vertical star count data in Fourier space, the data is transformed back into spatial data for our assessment of the mid-plane location.

For the present study, we have determined that a linear frequency cutoff value of $\Tilde{\nu}_{\rm lim} = 0.5 \ {\rm kpc}^{-1}$ is sufficient to remove the vertical waves, as such a cutoff value ensures we filter out (odd-parity) vertical structures on scales less than $\lambda_{\rm min} = {\Tilde{\nu}_{\rm lim}}^{-1} = 2$ kpc.  In other words, structure that ``fits" within the roughly 2 kpc vertical extent of the disk \citep[e.g.]{bovy2012milky} is filtered out, while any remaining odd-parity vertical structures are passed through the filter and therefore must be large scale, gradual variations.  To this end, we extend our study out to $|z| < z_{\ell}$, with $z_{\ell} = 2.0$ kpc.  While this choice of bounds will allow some halo stars into our calculation, we are interested in symmetry-breaking, {\it ``macro"} effects on the location of the mid-plane, and thus scales in excess of $\lambda_{\rm min} = 2$ kpc.

Finally, please note that we will use $\bar{z}$ and $\langle z \rangle$ in order to distinguish between the mid-plane location (average $z$ value) for the unfiltered and filtered data, respectively.

%
%
\section{Method and Data Selection} \label{sec:Method}

An assessment of the Milky Way mid-plane location is subject to observational biases in the Gaia data, extinction due to dust, as well as the superposition of various effects.  In order to circumvent the first of these issues, we employ a similar set of cuts to \citet{GHY20} and \citet{HGY20}, though we use Gaia DR3.  Namely, we select stars with $G \in [13, 18]$ mag, $G_{\rm BP} - G_{\rm RP} \in [0.6, 2.4]$ mag, within a few kpc of the Sun, slightly altering the color and magnitude ranges of \citet{GHY20}.

\begin{figure}[h!]
    \centering
    \includegraphics[scale=0.6]{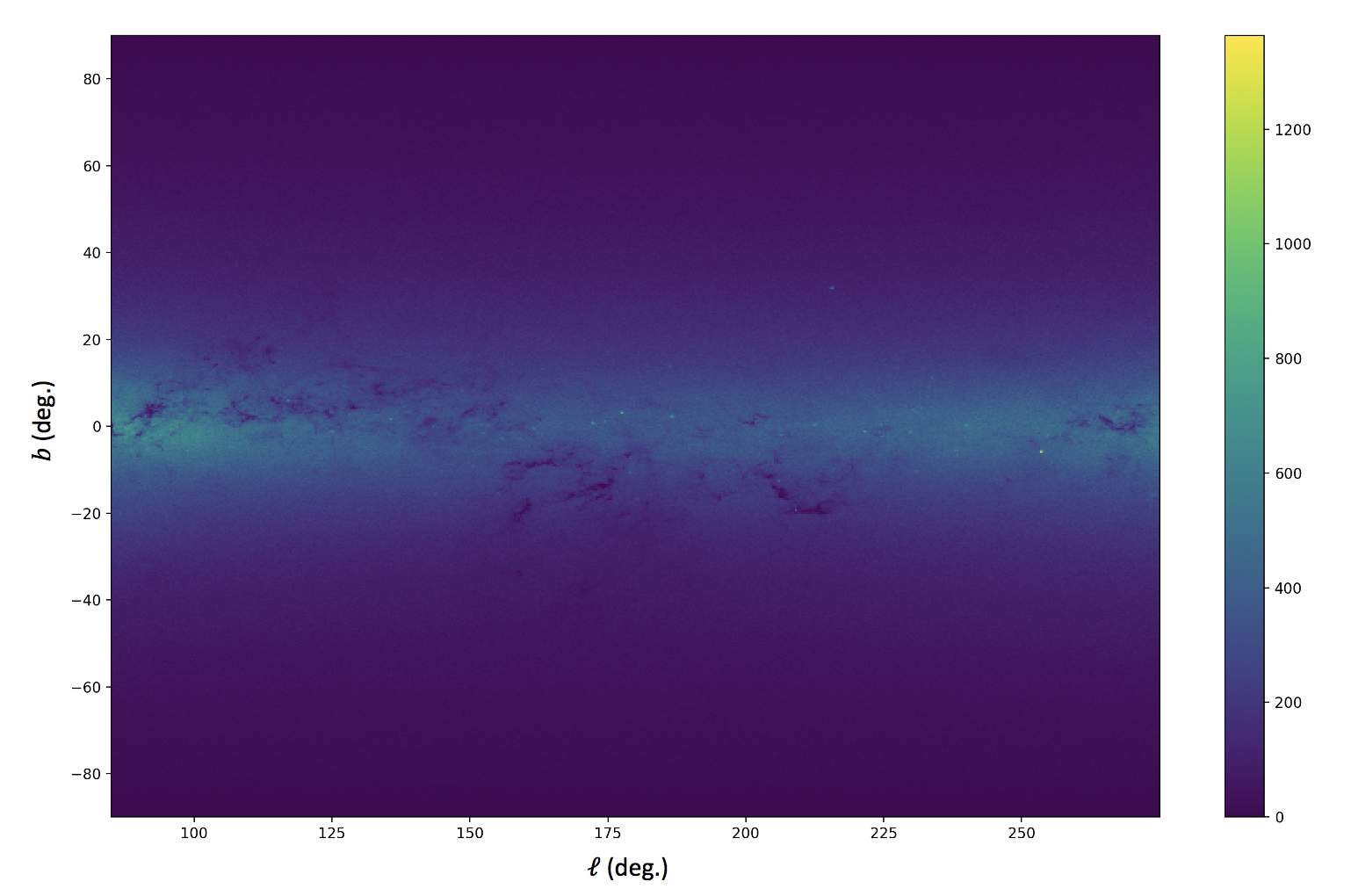}
    \caption{A longitude-latitude density plot illustrating the relatively dust-free line-of-sight towards $225^{\circ} < \ell < 245^{\circ}$, consistent with the maps of \citet{lallement2022updated}. We utilize this longitude window to extend the latitude range of \citet{GHY20} and \citet{HGY20} in order to study the mid-plane region.  }
    \label{fig:lbplot}
\end{figure}

However, the data set employed by \citet{GHY20} and \citet{HGY20} avoids the mid-plane region with latitude cuts of $|b| > 30^{\circ}$, whereas we are interested in including this highly-populated, low latitude region in order to determine variations in the height of the mid-plane.  This limits the lines-of-sight to those not severely impacted by dust, a relative rarity as seen in Fig.~\ref{fig:lbplot} as well as the studies of \citet{lallement2022updated}.  As such, we limit our study to heliocentric longitudes of $\ell \in [225^{\circ}, 245^{\circ}]$.

\begin{figure}[h!]
  \begin{center}
    \subfloat[]{\includegraphics[scale=0.56]{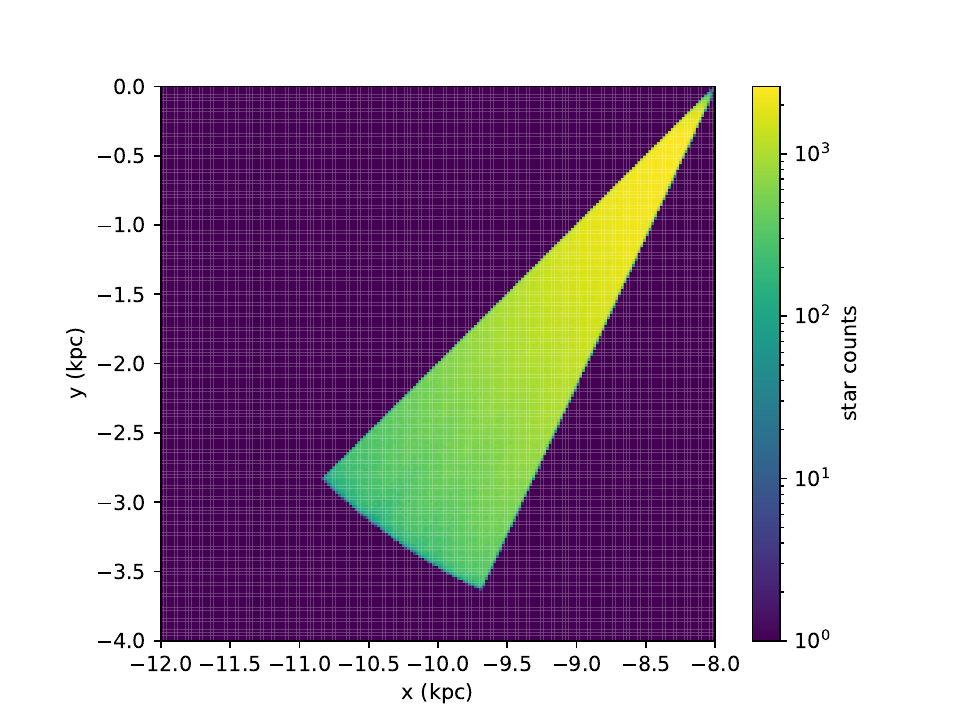}}
    \subfloat[]{\includegraphics[scale=0.56]{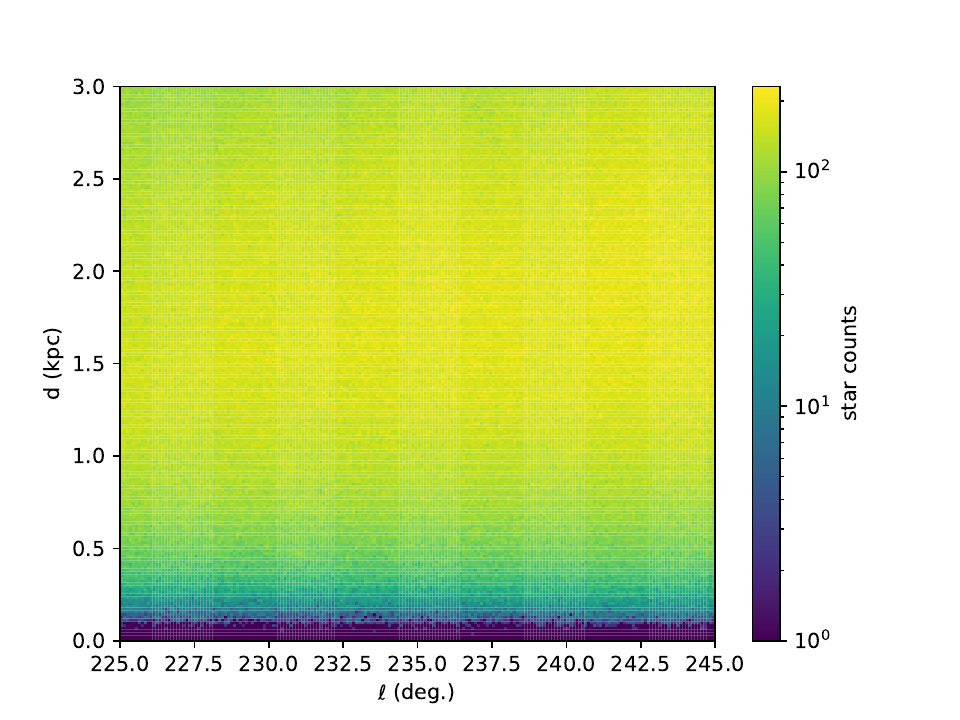}}
    \caption{
    (a) A Galactocentric $X-Y$ plot with the above selection along with a $|b| < 30^{\circ}$ cut to explore the impact of dust on the lower latitudes.  Note that this latitude cut is responsible for the artificial dearth of stars near the Solar location.  The absence of significant streaks pointing radially toward the Sun indicates that dust does not appreciably affect the completeness of this data set.  Some slight hints of dust occur beyond $d \approx 3$ kpc from the Sun, but this does not affect the conclusions in this paper.
    (b) A line-of-sight study along longitudes between $225^{\circ}$ and $245^{\circ}$ displaying star counts out to a 3 kpc distance from the Sun.  The absence of vertical streaks indicates minimal intrusion of dust, while the absence of other artefacts confirms that our extension of the data from \citet{GHY20} and \citet{HGY20} to lower latitudes does not introduce any incompletenesses.  
    }
  \label{fig:xyplot}
  \end{center}
\end{figure}

In order to confirm that the extension of the \citet{GHY20} and \citet{HGY20} data set to lower latitudes does not result in any newly introduced systematic errors, we reproduce figure 7a of \citet{HGY20} for the additional longitude selection motivated above.  Fig.~\ref{fig:xyplot}(a) displays this selected reproduction and indicates that no appreciable star count incompletenesses are introduced by extending the data to lower latitudes, and dust does not appear to occlude any lines of sight until about $d \approx 3$ kpc from the Sun, as illustrated in Fig.~\ref{fig:xyplot}(b).  Even then, this particular line of sight is {\it nearly} free of any occluding dust out to about 4 kpc from the Sun.

%
%
\section{Analysis} \label{sec:Analysis}

With our selection of data in hand, we divide the data into radial bins of width 100 pc.  For each radial bin, the vertical distribution of stars is histogrammed, Fourier Transformed, passed through the OLPF detailed in Eq.~\ref{Eq:OLPF} above, and Inverse Fourier Transformed back into position space.  The filtered data are then used to compute the average (filtered) position of the mid-plane, $\langle z \rangle$, for each radial bin.

\subsection{The Varied Morphology of the Vertical Waves}

\begin{figure}[h!]
  \begin{center}
    \subfloat[]{\includegraphics[scale=0.295]{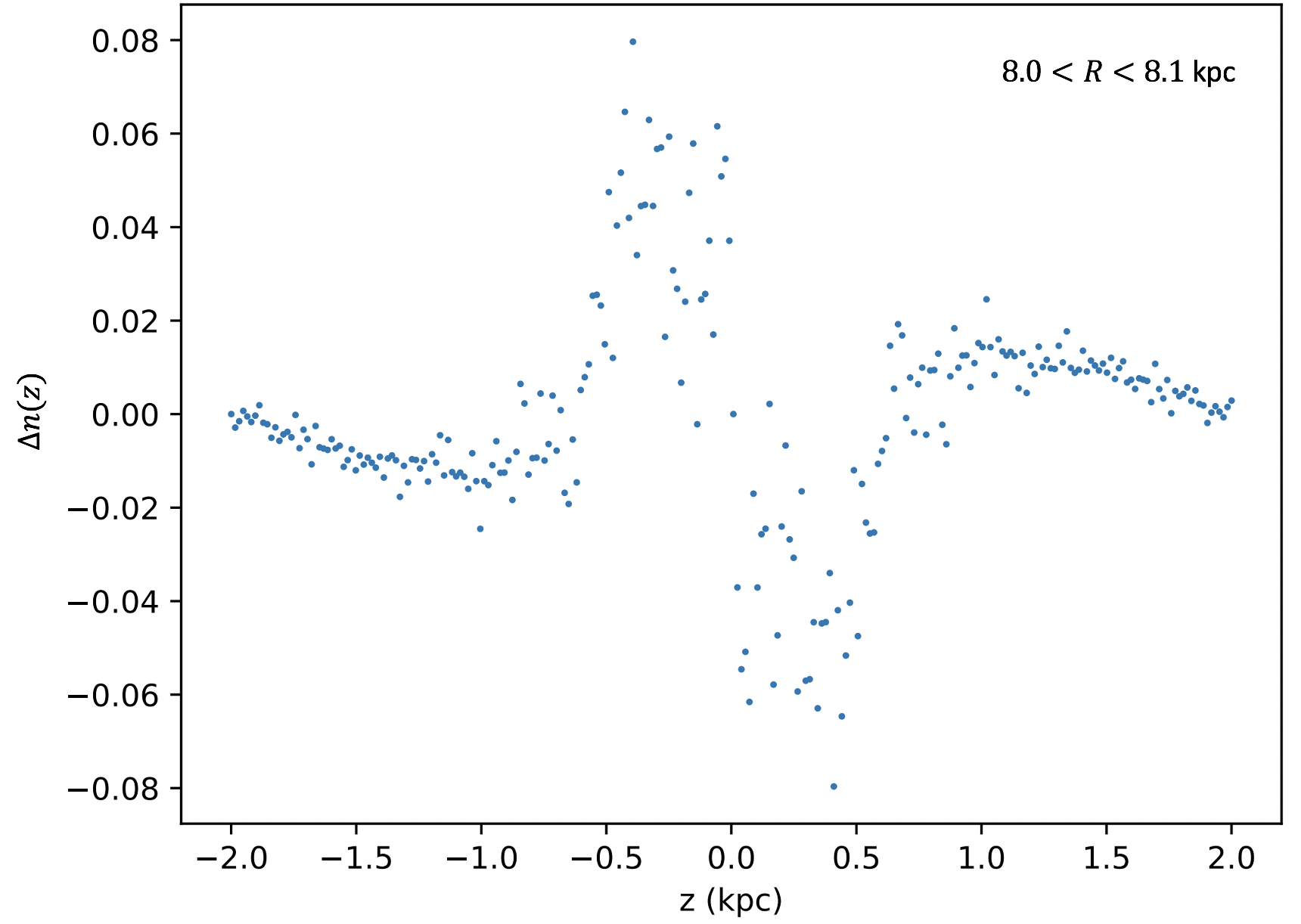}}
    \subfloat[]{\includegraphics[scale=0.295]{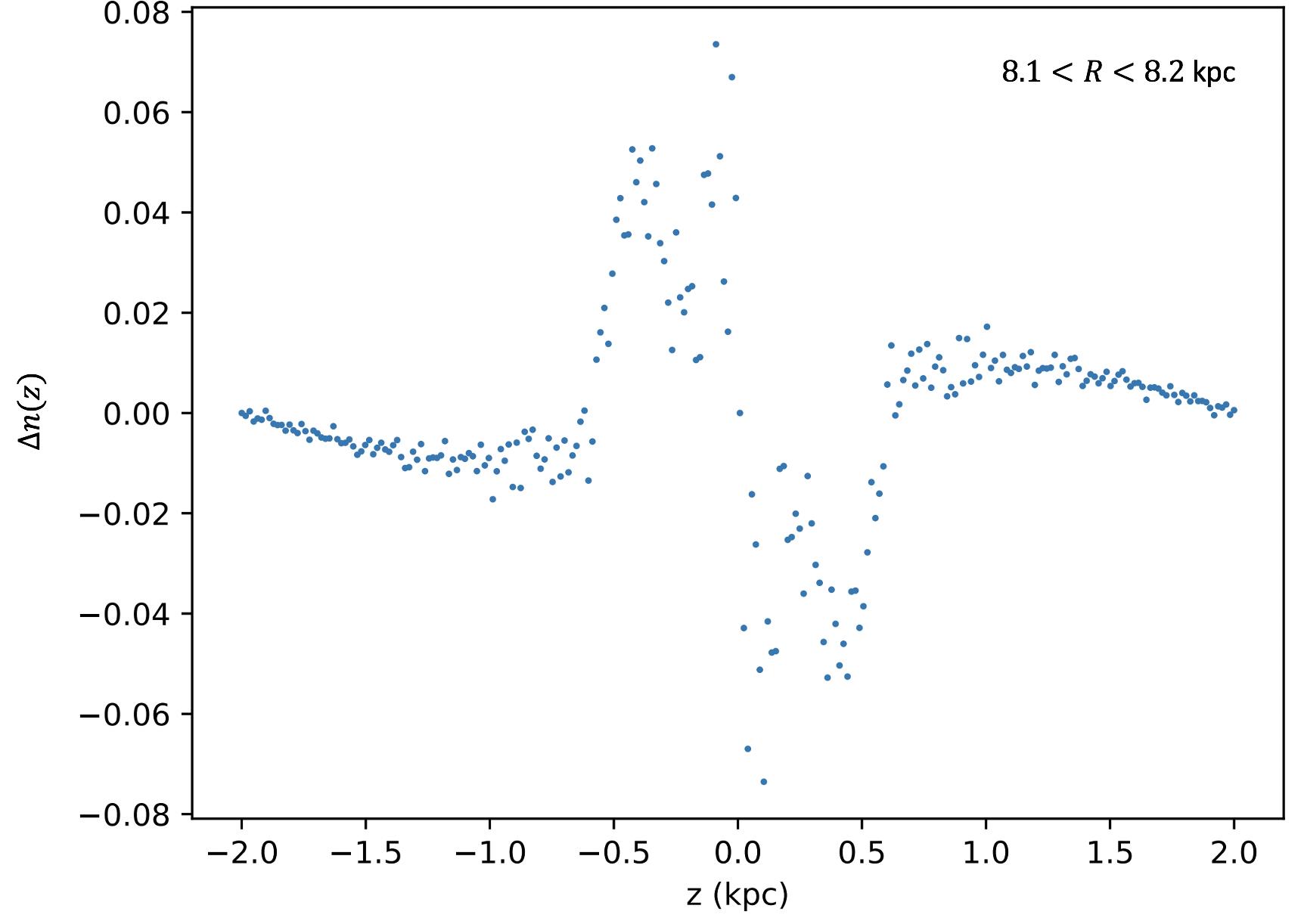}}

    \subfloat[]{\includegraphics[scale=0.295]{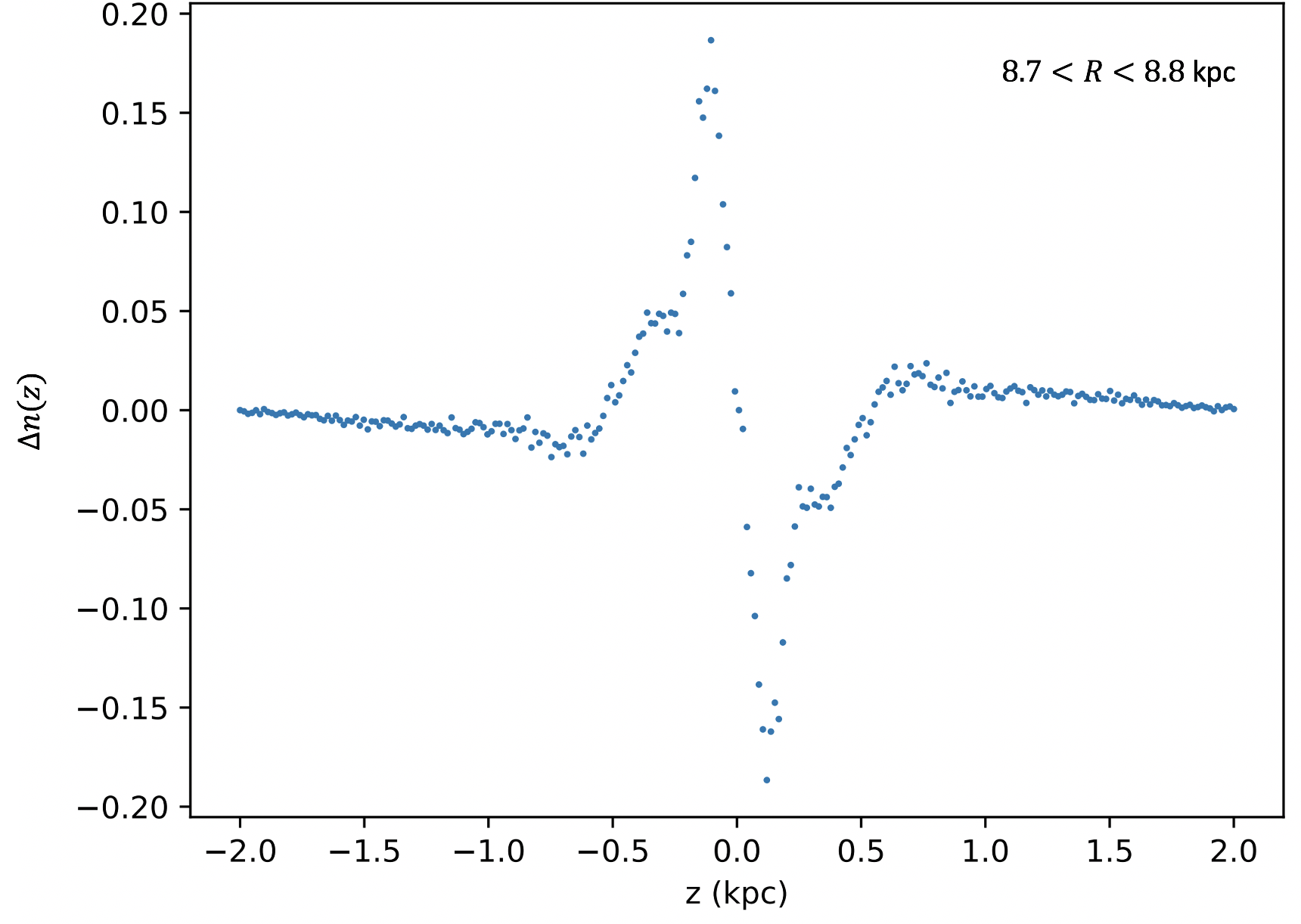}}
    \subfloat[]{\includegraphics[scale=0.295]{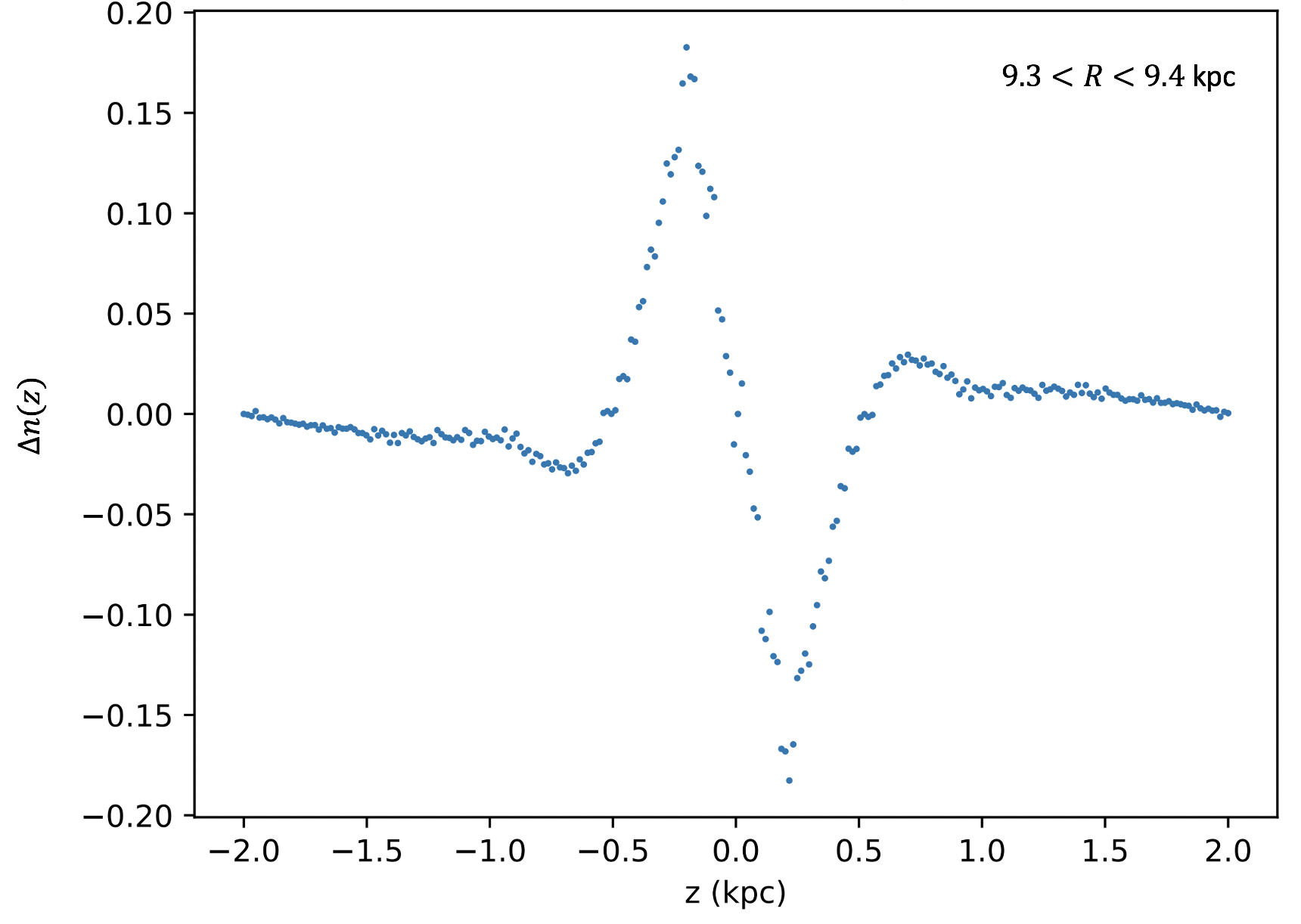}}
    \caption{
    The difference between the (normalized) raw histogram data and the (normalized) filtered histogram data, $\Delta n(z)$ for (a) $8.0 < R < 8.1$ kpc, (b) $8.1 < R < 8.2$ kpc, (c) $8.7 < R < 8.8$ kpc, (d) $9.3 < R < 9.4$ kpc, with error bars omitted for clarity.  The odd parity terms filtered out in our analysis of the mid-plane confirm the successful removal of small scale vertical structure and reveal the radially-changing morphology of the vertical waves of \citet{widrow2012galactoseismology}.  Please note that the narrow heliocentric longitude cuts ($225^\circ < \ell < 245^\circ$) limit the number of stars in the radial bins closest to the Sun, as illustrated in Fig.~\ref{fig:xyplot}(a).
    }
  \label{fig:filteredTerms}
  \end{center}
\end{figure}

To confirm that the OLPF acts as expected, we have explored the form of the odd terms that have been filtered out from each radial bin.  Namely, we isolate the filtered terms by subtracting the (normalized) filtered histogram data from the (normalized) raw histogram data: $\Delta n(z) = n_{\rm raw}(z) - n_{\rm filtered}(z)$, where each term was normalized to have a maximum of 1.
In doing so, we find that the vertical waves have indeed been correctly filtered out and, interestingly, that their morphology changes radially, as seen in Fig.~\ref{fig:filteredTerms}.

As a quick consistency check, we should expect $\Delta n(z)$ for the radial bin closest to the Sun\footnote{We round the solar radius to $R_0 = 8.0$ kpc for this analysis, though we note \citet{abuter2019geometric} for a more precise measurement.} 
to approximately match the vertical asymmetry found near the Sun in \citet{bennett2018vertical}, 
who explore a similar yet distinct volume of space.  Indeed, panels (a) and (b) of Fig.~\ref{fig:filteredTerms} 
depict the same extrema as \citet{bennett2018vertical} -- 
namely $|z| \sim 0.2$ kpc, $|z| \sim 0.4$ kpc, and $|z| \sim 0.7$ kpc, though limited statistics due to longitude cuts cloud the picture near the Sun.  
The Gaia data is also limited at very low latitudes ($|b| \sim 0^{\circ}$) due to crowded sight lines 
\citep[e.g.]{lindegren2018gaia}, which may be behind the sharp change around $z=0$ visible in some radial bins.

Although not a primary objective of our study, the changing morphology of the vertical waves is striking.  As one explores larger Galactocentric radii, $R$, as in panels (c) and (d) of Fig.~\ref{fig:filteredTerms}, the shape of the vertical waves changes, an effect noted in the two-point correlation studies of \citet{hinkel2023two} and perhaps related to the corrugated bending waves suggested by \citet{bland2021galactic}.

\subsection{A Radial Wave Extends Inward to the Solar Neighborhood}

\begin{figure}[h!]
  \begin{center}
    \subfloat[]{\includegraphics[scale=0.51]{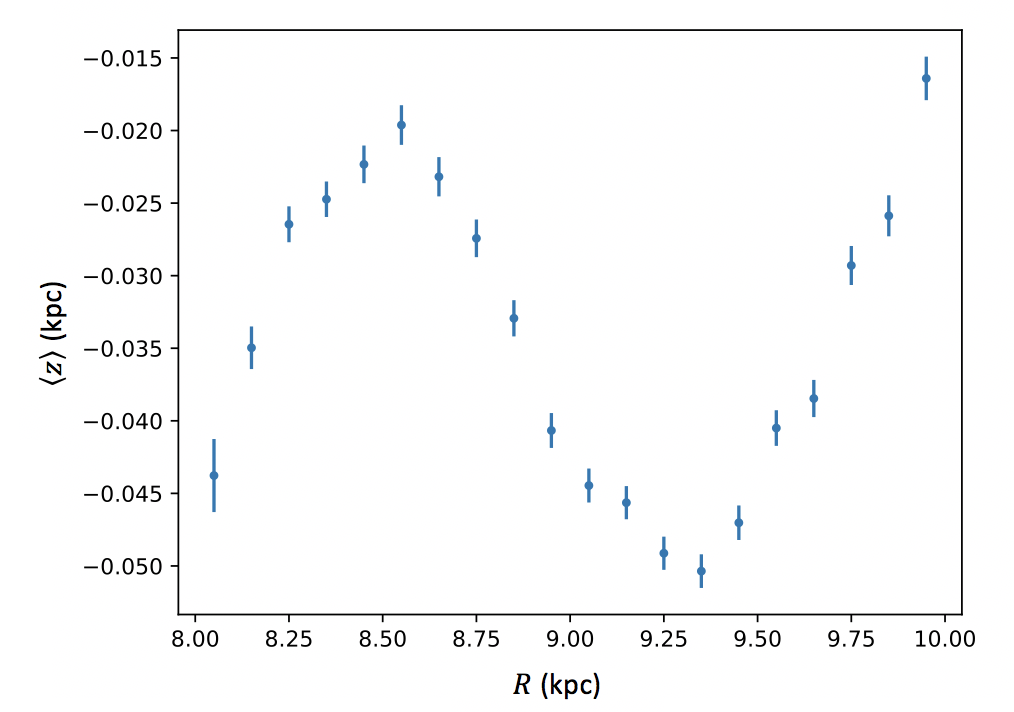}}
    \subfloat[]{\includegraphics[scale=0.51]{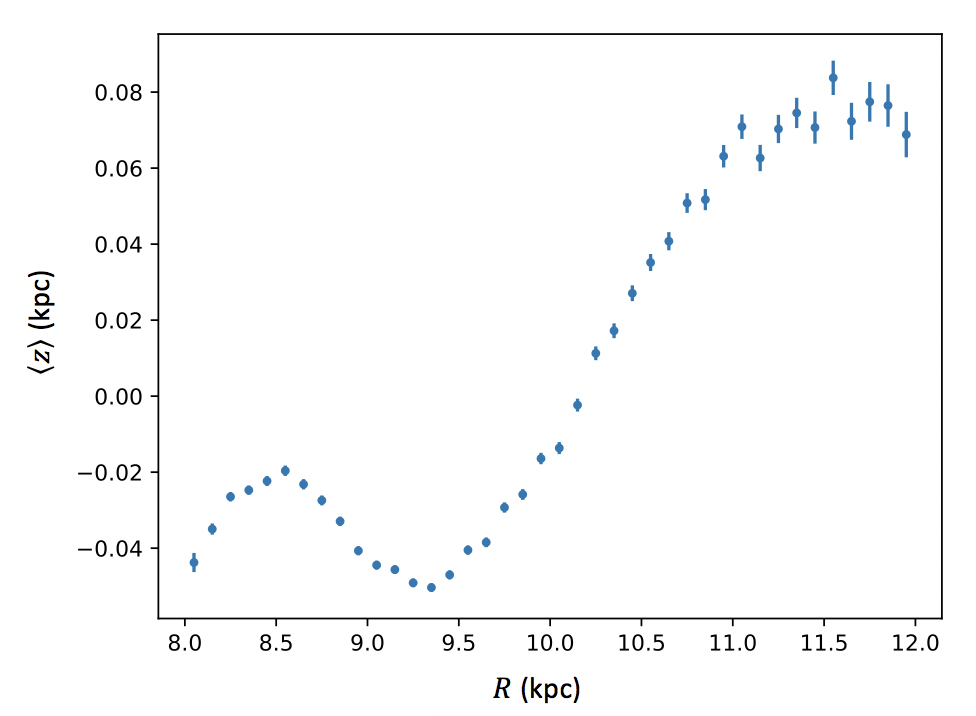}}

    \subfloat[]{\includegraphics[scale=0.51]{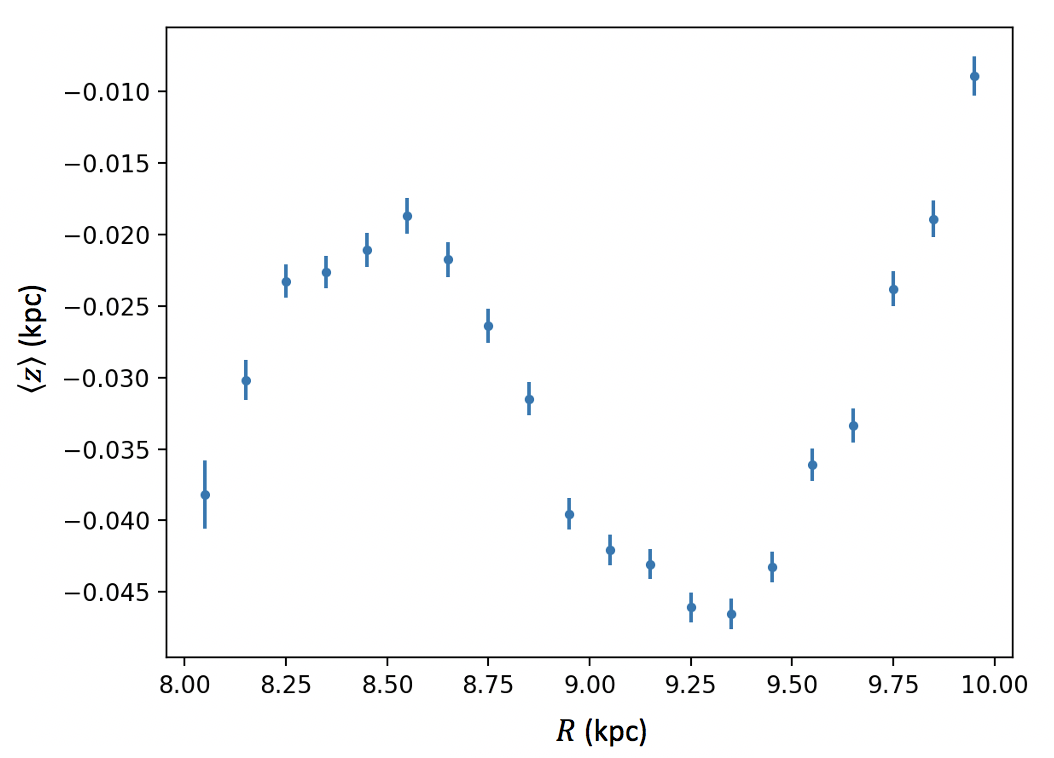}}
    \caption{
    (a) The post-OLPF, binned vertical star count data exhibits a radial wave of wavelength approximately 1.5 kpc and an amplitude of approximately 15 pc.  Note that the data has not been corrected for the solar offset, $z_{\odot}$, which, along with the OLPF results in the negative values of $z$.  
    (b) Extending the region examined radially outward, we find an additional maximum at 80 pc, which we believe is the 70 pc peak found in Figure 18 of \citet{xu2015rings}.  
    (c) The morphology of this radial wave remains the same for different choices of $z_{\ell}$.  (Pictured: $z_{\ell} = 1.5$ kpc.)
    }
  \label{fig:wave}
  \end{center}
\end{figure}

Upon computing the location of the mid-plane from the Odd, Low-pass Filtered data, we find evidence for a radial wave at galactocentric radii between $R = 8$ kpc and $R = 10$ kpc, firmly within the dust-free regions identified above.  The wave, pictured in Fig.~\ref{fig:wave}(a) exhibits an amplitude of approximately 15 pc and a wavelength of approximately 1.5 kpc. 
Indeed, the wave structure we have found appears to be a continuation of the radial wave discovered by \citet{xu2015rings}.  That is, the radial wave discovered by \citet{xu2015rings} warps the disk in the positive $z$-direction between about $R = 9$ kpc and $R = 12$ kpc, though we note that the vertical waves have not been filtered out as they are in the present work.  Additionally, \citet{xu2015rings} utilize SDSS DR8 data with isochrone-derived distances, which may also contribute to the slight discrepancies in the distances to the extrema of the radial wave. Indeed, we stretch our study to $R = 12$ kpc (Fig.~\ref{fig:wave}(b)) and find the larger, previously known maximum.

Please note that the data in Fig.~\ref{fig:wave}(a) is negative as it is offset by the Sun's position above the mid-plane, $z_{\odot}$.  Indeed, using the average of the Odd Low Pass Filtered data in Fig.~\ref{fig:wave}(a) as a small-scale-structure-corrected average mid-plane location over $8 < R < 10$ kpc, we determine that the local mid-plane lies at $\langle z \rangle = -34.2 \pm 0.3$ pc, implying $z_{\odot} = 34.2 \pm 0.3$ pc.  This wave-corrected estimate is roughly double the value found by \citet{karim2017revised}.  Indeed, the present finding is more consistent with an assessment made via Cepheids \citep{majaess2009characteristics} than studies employing number counts \citep{ferguson2017milky}, and is qualitatively consistent with the upward revision of the vertical asymmetry studies of \citet{bennett2018vertical}.  We confirm that the asymmetry of the vertical waves causes an over-density in the North in the highly-populated, low-$z$ region which, if not accounted for, will tend to artificially depress the value of $z_{\odot}$ when measured near the Sun.  

As others have noted \citep[e.g.]{ferguson2017milky}, the determination of $z_{\odot}$ via star counts depends on the direction one looks, as the concept of a mid-plane in a galactic disk is necessarily a local determination in a galaxy featuring substructure and warping.  Such a conclusion is reinforced here, as even upon filtering out the vertical waves and other smaller, purely vertical structures, the mid-plane of the Milky Way undulates vertically as one moves radially outward, consistent with a corrugation pattern \citep{bland2021galactic} and the slightly offset vertical waves found via a Two-Point Correlation function study \citep{hinkel2023two}.  That is, while our study endeavors to remove the vertical waves, there are still substantial differences in the mid-plane location along the radial direction, likely contributing (in part) to the various $z_{\odot}$ values found in the literature.  

Moreover, the morphology of the radial wave is robust against changes to our choice of $z_{\ell}$.  Indeed, the overall shape of the radial wave remains about the same as depicted in Fig.~\ref{fig:wave}(c), where we have changed $z_{\ell}$ to 1.5 kpc (c.f. Fig.~\ref{fig:wave}(a)).  There is, however, a small numerical change in the values of $\langle z \rangle$ with our choice of $z_{\ell}$, evident from the vertical axis scaling in panels (a) and (c) in Fig.~\ref{fig:wave} -- yielding a slight downwards shift with larger values of $z_{\ell}$.  This small shift may be due to the warping of the disk, which may disproportionately affect the thick disk stars \citep{widmark2022mapping}.

\subsection{A Heuristic Model for the Radial Wave}

In order to qualitatively explain the radial wave feature we have resolved, we have built a simple computational model via \texttt{python} that employs a Verlet integrator \citep{verlet1967computer}.  We use a Miyamoto-Nagai Disk profile \citep{miyamoto1975disk} 
as a static, background potential for the Milky Way disk, and embed small masses within this background potential\footnote{ The code for this simulation, as well as all of the code for the analysis in this manuscript can be found at \hyperlink{https://doi.org/10.5281/zenodo.10015988}{https://doi.org/10.5281/zenodo.10015988}}.  The small, embedded masses may move freely in the $z$ direction due to a perturbation, but are held fixed in the radial and azimuthal directions for simplicity.  

The system described above is not entirely unlike a wave on a massive string.  If we suppose the mass per unit length of the string increases monotonically in some direction, any wave pulse from the lower-mass end of the string will decrease in amplitude as it propagates toward the denser part of the string, as we expect energy to be conserved.   
Indeed, upon simulation, this is exactly the behavior we find (Fig.~\ref{fig:my_anim}).

\begin{figure}
    \centering
    \animategraphics[width=\textwidth, loop, autoplay]{5}
    {xz_plot-}
    {1}
    {33}
    \caption{An animation of a single wave pulse perturbation propagating from a low density region (right) to a region of higher density (left).  The initial amplitude of the wave is 15 pc, while the amplitude decays below 5 pc in the denser, interior region.  The simulation is meant to be qualitative only, and the limitations of such a simulation are discussed in the text.  The decaying amplitude qualitatively matches the radial wave we have studied in this paper.  
    While the simulation assumes the perturbation occurs at larger radii and penetrates in towards the center of the Galaxy, we do not discount the possibility of the perturbation starting at smaller radii and propagating outwards. }
    \label{fig:my_anim}
\end{figure}

This qualitatively matches the present results and the original finding of \citet{xu2015rings}.  To wit, we find evidence for a radial wave with larger amplitude at larger Galactic radii and a smaller amplitude at lower Galactic radii.  The feature is continuous, suggesting our finding is an extension of the discovery by \citet{xu2015rings}.  

To reiterate, this simulation is intended {\it only} to be qualitative and illustrative of the wave-like motion we are suggesting.  
The simulation is limited in a number of ways, including false, abrupt boundaries for the tracer populations, a static background potential inconsistent with non-steady-state behavior \citep{GHY20}, 
it ignores spiral structure and other structures, 
and treats the perturbed masses as a 1-D mass distribution 
confined to move only in the $z$-direction.  Indeed, the radial wave might be a bending wave of the disk itself, and so we do not claim to quantitatively model this wave.
Instead, we have employed this simulation purely to illustrate how a vertical perturbation to the disk can propagate radially, changing amplitude at various radii.  This later fact follows from conservation of energy, and although our heuristic simulation is limited, it nonetheless captures the variations in amplitude that we have found.

%
%
\section{Conclusions} \label{sec:Conclusions}

We have introduced a new tool for disentangling some of the myriad structures in the Milky Way via considerations of symmetry and scale.  Namely, we have developed and employed an Odd Low Pass Filter (OLPF) to remove the over- and under-densities of the vertical waves \citep{widrow2012galactoseismology}.  Upon filtering the vertical number count data along a relatively dust-free line of sight \citep{lallement2022updated}, we find evidence that the radial wave discovered by \citet{xu2015rings} penetrates further inward than previously understood -- to at least the Solar neighborhood.  

Indeed, we show further that the behavior of the radial wave qualitatively matches what one would expect for a wave pulse travelling along a medium with continuously changing mass density.  To wit: the amplitude of the wave is greater at larger Galactocentric radii and smaller in the denser, more interior regions, matching a simplistic, heuristic simulation we have built.  It is unclear at present which direction the radial wave is propagating, or even if it is perhaps some form of long-lived standing wave anchored between (e.g.) spiral arms.  

Refined radial velocity measurements from Gaia could potentially help to resolve the direction of propagation, revealing insights into the radial wave's origin.  If the wave propagates inward, is it a bending wave caused by an interaction with a subhalo?  If so, can we extrapolate backward in time to identify if the waves arise from the Sagittarius collision or perhaps a dark subhalo?  If the wave propagates radially outward, could it be due to bar buckling or some other internal process?

Intriguingly, the radial wave could also help to explain the existence of the Radcliffe Wave \citep{alves2020galactic}.  As \citet{swiggum2022radcliffe} suggest, a velocity difference of $10 - 20 \ \rm km \ s^{-1}$ between the spiral arm structure and the local gas and dust could explain the extent of the Radcliffe Wave.  However, \citet{swiggum2022radcliffe} note that the damped oscillatory behavior of the Radcliffe Wave remains unexplained.  If the spiral arms were themselves perturbed vertically by the radial wave, the gas and dust downstream from the spiral arms would have this vertical disturbance imprinted on it.  In this picture, as the gas and dust moves further from the region of the radial wave, the amplitude decays.

Assuming for the moment that the connection between the Radcliffe Wave and the radial wave is indeed real, the approximately 1 kpc wavelength of the Radcliffe Wave (Fig. 2: \citet{alves2020galactic}) along with the $10 - 20 \ \rm km \ s^{-1}$ velocity difference above implies a period of oscillation of about $50-100 \ \rm Myr$.  Given that the Radcliffe Wave exhibits at least one full wavelength of structure, this would then imply that the radial wave is at least $50-100 \ \rm Myr$ old.

Although tantalizing, the picture detailed above does not explain how the radial wave formed.  Moreover, our simulation employs a static background potential, which is at odds with the non-steady-state behavior of the Milky Way disk \citep{GHY20, HGY20}.  It would be interesting to model the system more fully in order to understand how a radially migrating bending wave might behave when interacting with the other substructures in the solar neighborhood.  We leave more detailed simulations to future work.

%
%
\section*{Acknowledgements}

The authors are grateful for funding from the Colorado College Faculty-Student Collaborative Grant
as well as research funding from the Natural Sciences Executive Committee at Colorado College.
A.H. also acknowledges Thomas More University for funding to finish and publish this work.  

This work has made use of data from the European Space Agency (ESA) mission
{\it Gaia} (\url{https://www.cosmos.esa.int/gaia}), processed by the {\it Gaia}
Data Processing and Analysis Consortium (DPAC,
\url{https://www.cosmos.esa.int/web/gaia/dpac/consortium}). Funding for the DPAC
has been provided by national institutions, in particular the institutions
participating in the {\it Gaia} Multilateral Agreement.

This work has also made use of the \texttt{numpy} \citep{harris2020array}, \texttt{sympy} \citep{10.7717/peerj-cs.103}, and \texttt{matplotlib} \citep{Hunter2007} packages.

\bibliography{zyBib, mybib}

\end{document}